\documentclass[aps, nofootinbib, twocolumn, showpacs,longbibliography]{revtex4-1}

\usepackage[CJKbookmarks, pdftex, bookmarksnumbered, bookmarksopen, colorlinks, citecolor=blue, linkcolor=blue]{hyperref}

\usepackage{amstext,amsmath,amssymb,amsfonts,bbm}
\usepackage[latin1]{inputenc}
\usepackage{graphicx}
\usepackage{epstopdf}
\usepackage{amsthm}
\usepackage{tocvsec2}
\usepackage{enumerate}
\usepackage{subfigure}
\usepackage{color}

\newcommand{\su}{{\scriptscriptstyle \cal S}}
\newcommand{\ta}{{\scriptscriptstyle \cal T}}
\newcommand{\BW}{{Bisognano--Wichmann }}

\def\be{\begin{equation}}
\def\ee{\end{equation}}
\def\bes{\begin{eqnarray}}
\def\ees{\end{eqnarray}}

\def\tr{{\rm tr}}

\begin{document}

\title{Thermally correlated states in Loop Quantum Gravity}

\author{Goffredo Chirco$^{1, 2}$, Carlo Rovelli$^{1, 2}$, Paola Ruggiero$^{3,4}$\\}

\affiliation{$^1$Aix Marseille Universit\'e, CNRS, CPT, UMR 7332, 13288 Marseille, France.\\
$^2$Universit\'e de Toulon, CNRS, CPT, UMR 7332, 83957 La Garde, France.\\
$^3$Dipartimento di Fisica dell'Universit\`a di Pisa, Largo Pontecorvo 3, I-56127 Pisa, Italy.\\
$^4$SISSA, Via Bonomea 265, 34136 Trieste, Italy.
}

\date{\today}

\begin{abstract} 

\noindent 
We study a class of loop-quantum-gravity states characterized by (ultra-local) thermal correlations that reproduce some features of the ultraviolet structure of the perturbative quantum field theory vacuum. In particular, they satisfy an analog of the Bisognano-Wichmann theorem. 
These states are peaked on the intrinsic geometry and admit a semiclassical interpretation. We study how the correlations extend on the spin network beyond the ultra local limit. 

\end{abstract}

\maketitle

\section{Introduction}
The \BW theorem \cite{Bisognano:1976za} states that the restriction of the vacuum of a Lorentz-invariant quantum field theory to the algebra of field operators with support on the Rindler wedge $x\!>\!|t|$ is a KMS (Kubo-Martin-Schwinger) state, that is, a thermal equilibrium state \cite{Haag:1967sg}, with inverse temperature $2\pi$, with respect to the flow generated by the boost operator $K$ in the $x$ direction \cite{Haag:1992hx}. That is, it is described by the density matrix
\begin{equation} \label{minktherm}
\rho \propto e^{- 2 \pi K} .  
\end{equation}
This fact is at the root of the thermal aspects of quantum field theory, such as the Unruh effect \cite{Unruh:1976db}.  On a curved spacetime, quantum fields mimic flat space properties locally, and (a local version of) \eqref{minktherm} can be argued to underpin the thermal properties of black holes \cite{Unruh:1976db,Bombelli1986,Srdnicki1993,Jacobson:2003wv,Solodukhin2011,Bianchi2012,Bianchi2012b}.\footnote{Hawking's black hole temperature is precisely equal to the Unruh temperature observed by a stationary observer near the horizon, red-shifted from this observer's location to infinity; see \cite{Frodden:2011eb}.}  

This effect derives from the quantum correlations in the field.  This is particularly clear by considering states at fixed time. Eq.~\eqref{minktherm} can be obtained --at least formally-- by tracing the state on the $t=0$ surface over the degrees of freedom with support on $x<0$.  The resulting state is not pure because of the field correlations across $x=0$. In general, we say that a state on a 3d spatial surface $\Sigma$ has the \BW property if in any sufficient small patch of $\Sigma$ a version of \eqref{minktherm} holds locally for any 2d surface $S$, after tracing over the degrees of freedom on one side of $S$.  (More precision below.) This property captures aspects of the field's local correlations.  

This is of interest in quantum gravity for the following reason.  The full background-independent nonperturbative theory must include states yielding conventional physics at low energy, including quantum field correlations. But the ultraviolet structure of these correlations which characterises theories defined on a background geometry
\be
  \langle 0 |\phi(x)\phi(y)| 0 \rangle \sim \frac1{|x-y|^2}, 
  \label{shortscale}
\ee
does not remain true in a quantum gravity theory (such as loop quantum gravity) where the Planck scale is a physical cut-off and there is no background metric defining the distance on the right  hand side of this equation. Thus, \eqref{shortscale} is not useful for characterising semiclassical states. On the other hand, as we shall see, \eqref{minktherm} makes sense naturally in the theory. And it better hold true for semiclassical states, for these to yield the expected low energy phenomenology.\footnote{Entanglement entropy due to short-scale quantum correlations has been studied in loop quantum gravity, especially in the context of black holes thermodynamics \cite{Ashtekar:1997yu,Bianchi:2012ui,Frodden:2011zz,BarberoG.:2012ae,Ghosh2013}.}  Here we explore the possibility of using  \eqref{minktherm} as a (partial) characterisation of ``good" semiclassical states in quantum gravity. 

A similar suggestion has been recently put forward by two papers. In \cite{Bianchi2012b}, Bianchi and Myers have suggested Bisognano--Wichmann-like correlations to characterize semiclassical states in any nonperturbative quantum theory of spacetime. The smooth structure of space-time geometry at the classical level may be intimately related to the structure of correlations of the quantum gravitational state. A similar perspective has received attention in string theory, in the context of the gauge/gravity duality, where the entanglement of the boundary gauge field degrees of freedom has been associated to the connectivity of the bulk space-time dual \cite{VanRaamsdonk2010,VanRaamsdonk2010b,Czech2012a}.  In \cite{Chirco2014}, the \BW property has been suggested as a possible replacement, in a background independent context, of the Hadamard condition that characterises the ``good" states in quantum field theory on curved space.  

In loop quantum gravity,  semiclassical states have been studied extensively \cite{Thiemann:2002vj,Livine:2007mr,Bianchi:2009ky,Bianchi:2010gc,Ashtekar:2013hs}. Today we know how to write states where the \emph{expectation value} of the the gravitational field appropriately matches a given smooth geometry.  However, little is known so far about states where also the \emph{fluctuations} of the gravitational field, and especially the nonlocal correlations, match the ones of conventional field theory.  Here we construct and study states with a Bisognano--Wichmann-like property, as a step in this direction.  

Notice that the main hypotheses of the \BW theorem are positivity of energy and Lorentz invariance. The last is a dynamical property in the sense that a boost generates the change of a state from a given (spacelike) plane to a boosted one.  Therefore the \BW property captures aspects of a state's evolution. As we will see, this is reflected in the states we define below: their definition depends on the (covariant \cite{Freidel:2007py,Engle:2007wy,Kaminski:2009fm,Rovelli2011c,Rovelli,Ashtekar:2013hs}) definition of the loop quantum dynamics. Therefore they can also be viewed as a step towards fully physical dynamical quantum gravity states.
 
Section \ref{math} recalls the covariant definition of loop quantum gravity states.  In section \ref{hadamard} we define the thermally correlated link state, the fundamental brick of our construction. In Section \ref{semi} we study the semiclassical properties of this state. In Section \ref{sntherm} we construct the thermally correlated $SU(2)$ spin network state. Section \ref{propa} shows how local correlations `propagate' along the spin network. %Eventually Section \ref{enta} provides an explicit expression for the non-local contributions to entanglement in the thermally correlated spin network.  
Results are summarised and discussed in the last section.

\section{Lorentz covariant LQG states}\label{math}

We start introducing the conventional loop quantum gravity space state, but using the $SL(2,\mathbb{C})$ covariant language \cite{Alexandrov:2002br,Dupuis:2010bh,Rovelli:2010ed} adapted for what follows.

\begin{figure}[t]
\includegraphics[width=2.5 in]{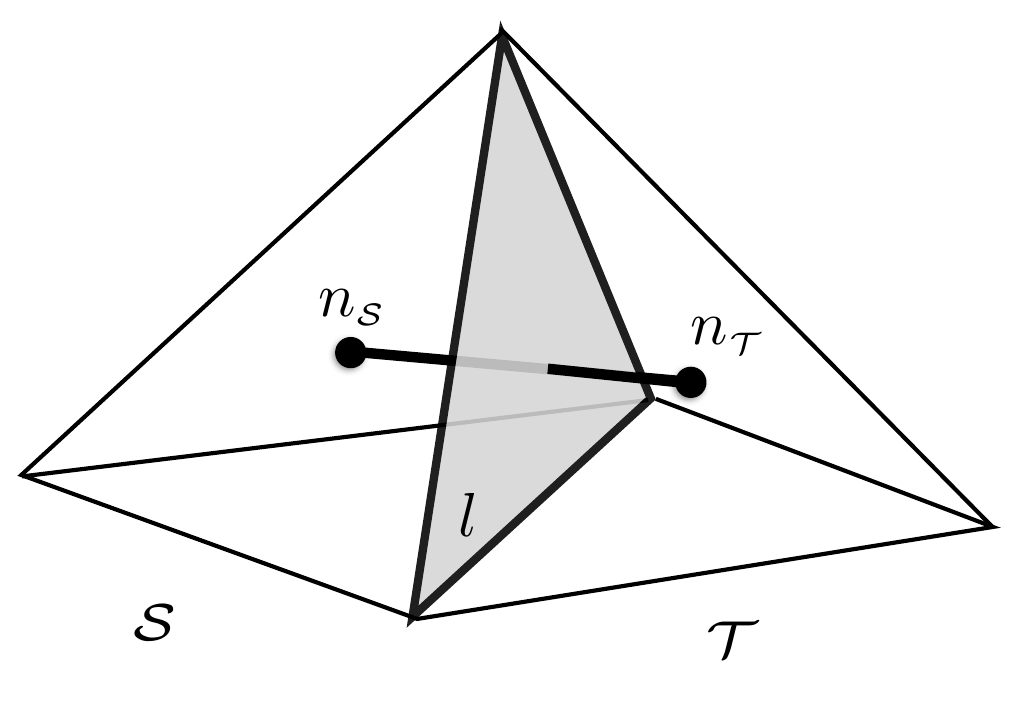}
\caption{A facet $l$ separating two cells (a source cell $\su$ and a target cell $\ta$) punctured by the link $l$ that joins the two corresponding nodes.}\label{surface}
\end{figure}

Consider an oriented three dimensional space-like hypersurface $\Sigma$ embedded in a four dimensional space-time manifold $\mathcal{M}$. Fix an oriented cellular decomposition of $\Sigma$.  We call $n$ (for ``node") the cells and $l$ (for ``link") the facets separating two adjacent cells. Fix a dual graph $\Gamma$, with a node $n$ in each cell and a link $l$ connecting the nodes of neighbouring cells. We use the same notation, $n$ and $l$ for the nodes and links of the graph and the dual cells and facets of the triangulation.  Each link $l$ is oriented: we call $n_\su$ (for ``source") and $n_\ta$ (for ``target") its initial and final nodes. The corresponding facet is equally oriented and separates a ``source" cell $\scriptstyle \cal S$ from a ``target" cell ${\scriptstyle \cal T}$. See Figure \ref{surface}.

Consider the spin connection $\omega$ of the Cartan formulation of general relativity, restricted to $\Sigma$. This describes aspects of the gravitational field on $\Sigma$. A quantum state can be expressed as a functional of $\omega$. In particular, a quantum state on $\Gamma$ is defined to be a (cylindrical\footnote{A cylindrical function on an infinite dimensional space is a function depending on only on a finite number of coordinates on this space.}) function $\Psi[\omega] = \psi(g_l[\omega])$ of the holonomy $g_l[\omega]\in SL(2,\mathbb{C})$ of the spin connection along the $L$ links $l$ of the graph. These states, we assume, can be expanded in matrix elements of unitary representations of $SL(2,\mathbb{C})$. 

The $SL(2,\mathbb{C})$ generators $J_{l}^{IJ}=-J_l^{JI}, I,J=0,...,3$ associated to each oriented link $l$ play the role of the basic observables of the theory. They are the quantum operators representing the momentum conjugate to the spin connection, which on $\Sigma$ is proportional to the Plebanski two-form $e^I\wedge e^J$, where $e^I$ is Cartan's tetrad one-form. More precisely, they are determined by the flux of this quantity across the facet $l$ 
\be
    J_{l,\su}^{IJ}\sim \int_l e^I\wedge e^J,
\ee
parallel transported to the source node $n_\su$ of $l$. The operator $J_{l,\su}^{IJ}$ acts on a function of $g_l$ as the left-invariant vector field.  It is important for what follows to observe that the \emph{right}-invariant vector $\tilde J_{l,\ta}^{IJ}$, related to $J_{l,\su}^{IJ}$ by the transformation defined by $g_l$ (in the adjoint representation), is a distinct operator. It represents the same flux, but parallel transported to the \emph{target} node $n_\ta$ of the link $l$, namely in the frame of the adjacent cell.  The relation between the two operators depends on the spin connection along the link. 

It is convenient to pick the time gauge, which ties the normal to $\Sigma$ to a direction $t$ in the internal Minkowski space.  Then $J_l^{IJ}$ splits into rotation generators $\vec L$ and boost generators $\vec K$. It is easy to see that (in a locally flat context \cite{Dittrich2014}) the first is a vector normal to the facet $l$, with length proportional to the area of the facet \cite{Rovelli}.  

The unitary representations of $SL(2,\mathbb{C})$ are labelled by a positive real number $p$ and non negative half-integer $k$ \cite{Ruhl:1970fk}. At each node, the vector $t$ determines a subgroup $SU(2) \subset SL(2,\mathbb{C})$ that leaves it invariant. The Hilbert space ${\cal{H}}_{(p,k)}$ that carries the $(p,k)$ representation decomposes into irreducible representations of the subgroup as follows
\be
{\cal{H}}_{(p,k)}=\oplus_{j=k}^{\infty}{\cal{H}}_j,
\ee
where ${\cal{H}}_j$ is the (finite dimensional) $SU(2)$ representation of spin $j$.  Therefore ${\cal{H}}_{(p,k)}$ admits a basis $|(p,k);j,m\rangle$, called the canonical basis, obtained by diagonalizing the total angular momentum $L^2$ and the $L_z=\vec{L}\cdot \vec{z}$ component of the $SU(2)$ subgroup. The states 
\be
       \Psi_{p_lk_lj_lm_lj'_lm'_l}[\omega]=\otimes_l \ D^{(p_l,k_l)}_{j_lm_l,j_l'm_l'}(g_l[\omega]), 
\ee
where $D^{(p_l,k_l)}$ are the representations matrices of the $SL(2,\mathbb{C})$ unitary representations, span the space of the states on $\Gamma$.

Within ${\cal{H}}_{(p,k)}$ the \emph{physical subspace} of the theory is determined (in a given Lorentz frame) by the linear simplicity condition 
\be
\vec{K}=\gamma \vec{L}
\label{linear}
\ee
satisfied, in general relativity, by the momentum conjugate to the spin connection.  Here $\gamma \in \mathbb{R}^+$ is the Barbero-Immirzi parameter \cite{Engle:2007wy,Rovelli}. This relation determines a restriction on the set of the unitary representations and picks a subspace within each representation. Indeed, the relation \eqref{linear} is weakly (in matrix elements) true \cite{Ding:2009jq} when restricted to states of the form
\be
|p,k;j,m\rangle = |\gamma(j+1),j;j,m\rangle. 
\ee
Accordingly, the physical subspace is formed by the states of the form
\be
       \Psi_{j_lm_lm'_l}[\omega]= \otimes_l \ D^{(\gamma(j_l+1),j_l)}_{j_lm_l,j_lm_l'}(g_l[\omega]) .
\label{questo}
\ee
This state space is naturally isomorphic to the space $L_2[SU(2)]^L$, the conventional (non gauge invariant) Hilbert space of loop quantum gravity on the graph $\Gamma$.  The isomorphism maps \eqref{questo} into 
\be
       \psi(h_l)= \otimes_l \ D^{(j)}_{m_l,m_l'}(h_l) 
\ee
where $h_l\in SU(2)$ and $D^{j} (h)$ are Wigner matrices, and is determined by the injection 
\bes
Y_\gamma: \ \ \ & &{\cal H}_{j} \ \ \ \to{{\cal H}_{(\gamma (j+1),j)}} \\ \nonumber
&& |j,m \rangle \mapsto |(\gamma (j+1),j); j,m\rangle.  \nonumber
\ees 
$L_2[SU(2)]^L$ is a Hilbert space and this isomorphism endows the physical state space with the scalar product needed to define a quantum theory. 

Let us now see how local gauge invariance affects this construction. In the Cartan formulation, general relativity is invariant under local $SL(2,\mathbb{C})$ gauge transformations. Of these, only the Lorentz transformation $\Lambda_n$ at the nodes $n$ of $\Gamma$ affect the states on $\Gamma$ (because only these affect the holonomies $g_l[\omega]$).  Consider first gauge transformations where $\Lambda_n$ are rotations. These do not affect the local frame at each node, and transform physical states into themselves.   The states invariant under these transformation are the well known spin network states
\begin{equation}
\psi(h_l)=  \bigotimes_{l} D^{j_l} (h_l) \cdot \bigotimes_{n} \iota_n
 \label{stato}
\end{equation}
where $\iota_n$ is an $SU(2)$ intertwiner at the node $n$ and the contraction is determined by the structure of the graph. These gauge invariant states form the Hilbert space $\mathcal{H}_{\Gamma}$
\begin{equation*}
\mathcal{H}_{\Gamma}=L^2[SU(2)^L/SU(2)^N ]
\end{equation*}
the standard loop quantum gravity state space on a graph.  The states in this space have a direct interpretation as quantum geometry of the spatial section $\Sigma$ of space-time. 

More interesting are the Lorentz transformation that are not rotations. These act on the $SL(2,\mathbb{C})$ states, changing (rotating) the class of physical states.  Say $t$ is a vector in the Minkowski representation, left invariant by $SU(2)$;  a generic Lorentz transformation boosts $t$ into $\Lambda t$, which stabilises a different $SU(2)$ subgroup, which in turn defines a different class of physical states.  Therefore the spin network formalism is invariant under local rotations but is covariant under boosts. See \cite{Rovelli:2010ed}  for a full discussion. 

We are interested in the structure of correlations of these states. In other words. we are interested in the way different regions of a spin network can be correlated to one another.

\section{Thermal link states} \label{hadamard}

\subsection{\BW property on a single link}

Let us begin by focusing on a single link, and disregarding, for now, gauge invariance.  The states on a single link are given by functions $\psi(g)$ on $SL(2, \mathbb{C})$ satisfying the simplicity constraint, that is, linear combinations of the states of the form 
\be
       \Psi_{jmm'}(g)\equiv \langle g|jmm'\rangle =  D^{(\gamma(j+1),j)}_{jm,jm'}(g). 
\ee
The operator $\vec L_\su$ acts on this state as the generator of rotations on the first index
\be
      \langle jmm'| \vec L_\su| jm''m'''\rangle =  \vec \tau^{j}_{m m''} 
\ee
where $\vec \tau^j$ is the generator of rotations in the spin $j$ representation of $SU(2)$ and summation over related indices is understood. The operator $\vec L_\ta$ acts on this state as the generator of rotations on the second index
\be
      \langle jmm'| {\vec L_\ta} | jm''m'''\rangle =  \vec \tau^{j}_{m'm'''}.  
\ee
The two boost generators, $\vec K_\su$ and ${\vec K_\ta}$ restricted to this space, have the same matrix elements, multiplied by $\gamma$. This set of operators splits naturally into two groups: $\vec L_\su$ and $\vec K_\su$ act on the first magnetic index and represent observables living on the cell on the source side of the facet; while $\vec L_\ta$ and $\vec K_\ta$ act on the second magnetic index and represent operators living on the cell on the target side of the facet. Recall that they represent quantities parallel transported to two different nodes, and their difference measures the connection along the link. We can therefore split the observables into two groups, associated to the two cells on opposite sides of the facet $l$.   

All operators considered here are diagonal in $j$ (the boost operator mixes different $j$ sectors of the same $SL(2, \mathbb{C})$ irreducible, but not different  $SL(2, \mathbb{C})$ irreducibles, of course; also, states with different $j$ belong to different irreducibles, because $k=j$). It is therefore convenient to work at fixed quantum number $j$, namely on a $L^2$ eigenspace (clearly $|\vec L_\su|^2=|\vec L_\ta|^2$). This space has the structure
\be
          {\cal H}={\cal H}^\su_j\otimes{\cal H}^\ta_j.
\ee
Given a state in this subspace, we can trace on one factor and define a density matrix over the other. Explicitly, tracing on the target factor, a state of the form 
\be
     |\psi\rangle =\sum_{mn} c_{mn} | jmn\rangle 
\ee
gives the density matrix
\be
     \rho =Tr_\ta  | \psi\rangle\langle \psi| \equiv  \sum_n \overline{c_{nm}}c_{nm'}|j,m\rangle \langle j,m'|
\label{traccia}
\ee
on ${\cal H}^{\su}_j$. Since the restriction of $\vec K$ to ${\cal H}_j$ is given by $\gamma \vec L$, because of the simplicity conditions, we can define the density matrix 
\be
     e^{-2\pi \vec K\cdot \vec z} = \sum_m e^{-2\pi \gamma m} |j,m\rangle \langle j,m|.
\ee
where here $|j,m\rangle$ is a basis of eigenstates of $\vec L\cdot \vec z$.  We now have the language for the following definition.  We say that a link state $\psi$ with spin $j$ has the \textit{\BW property} if there is a $\vec n$ such that 
\be
\tr_\ta  [| \psi\rangle\langle \psi|] = e^{-2\pi \vec K_\su \cdot \vec n}.
\label{BW}
\ee
\emph{and} there is a  $\vec n'$ such that 
\be
\tr_\su  [| \psi\rangle\langle \psi| ]= e^{-2\pi \vec K_\ta \cdot \vec n'}.
\label{BWW}
\ee
Armed with this definition, let us now see what are the states with this property. 

\subsection{States}

We want to find a class of states $\{ |\psi \rangle \}$ satisfying (\ref{BW}) and (\ref{BWW}). 

For a given $\vec{z}$, we set $\vec{n} = \vec{n}' \equiv \vec{z}$. Sandwiching (\ref{BW}) between eigenstates of $\vec K_{\su} \cdot \vec z$ gives 
\be
\langle jm\, | \,\tr_\ta [|\psi \rangle \langle \psi |]\,|  j m' \rangle =e^{-2 \gamma \pi m} \delta_{mm'} 	
 \label{condition_kl}
\ee
Using \eqref{traccia}, this reads
\be
 \sum_n \overline{c_{nm}}c_{nm'}=e^{-2 \gamma \pi m}\ \delta_{mm'} .	
\ee
Let $\Lambda$ be the diagonal matrix with entries $e^{-\gamma \pi m}$ and $c$ be the matrix with matrix elements $c_{nm}$. Then the last equation can be written in the form
	\be
	c c^\dagger  = \Lambda \Lambda^{\dagger}
	\ee
	or equivalently
	\begin{equation*}
(c\,\Lambda^{-1} ) (c\,\Lambda^{-1} )^{\dagger} = \mathbb{I},  
	\end{equation*}
which is solved for any unitary matrix $U$ by 
	\begin{equation}
	c = \Lambda U  
	\end{equation}
In components, our coefficients read
	\begin{equation} \label{U}
	c_{mn} = e^{-\pi \gamma m}\, U_{mn}
	\end{equation}
Moreover, recall that the definition of the \BW property demands the state to be thermal when traced on either side. Repeating the above derivation with source and target swapped yields
	\begin{equation} \label{V}
	c_{mn} = V_{mn}e^{-\pi \gamma n}\, 
	\end{equation}
with $V_{mn}$ also a unitary matrix.	

Now, for two generic directions $\vec{n}$, $\vec{n}'$, it is easy to show that equations (\ref{U}) and (\ref{V}) generalize to
\begin{eqnarray} \label{U_n}
c_{mn} &=& \sum_k D(\vec{n})_{mk}e^{-\pi \gamma k}U_{kn}\,  \\ 
 c_{mn} &=& \sum_k V_{mk}e^{-\pi \gamma k}D^{\dagger}(\vec{n}')_{kn}\,  \label{V_n}
\end{eqnarray}
where $D(\vec{n})$ and $D(\vec{n}')$ are the Wigner matrices (in the representation $j$) corrisponding to the $SU(2)$ elements rotating the $\vec{z}$ into the $\vec{n}$ and the $\vec{n}'$ axes, respectively.	

A wide class of states satisfying both (\ref{U_n}) and (\ref{V_n}) is given by
%\textcolor{red}{And combining these two requirements it is easy to see that $U$ and $V$ must be $SU(2)$ representation matrices.  	
%Thus, the general state with spin $j$ satisfying the \BW property has the form 
	\begin{eqnarray}
	|\psi\rangle &=& \sum_{mnl} D(\vec{n})_{lm} e^{-\pi \gamma m}  D^{\dagger} (\vec{n}')_{mn} |j, l, n \rangle
	\end{eqnarray}
Note that the effect of the Wigner matrices on the basis states $ |j, l , n \rangle = |j, l  \rangle\otimes  |j, n \rangle^{\dagger}$ is simply to transform the $L_z$ eigenbasis $ |j, l \rangle$ into the eigenbasis  $ |j, l \rangle_{\vec n}$ of $\vec L\cdot \vec n$ for a generic vector $\vec n$. Therefore this class of (spin $j$) states labelled by two arbitrary vectors and the $SU(2)$ representation $j$, that satisfy the \BW property, has the compelling from 
	\begin{eqnarray}
	|\psi_{j\vec n \vec n'}\rangle &=& \sum_{m} e^{-\pi \gamma m}  |j, m \rangle_{\scriptscriptstyle \vec n} \otimes | j, m \rangle_{\scriptscriptstyle \vec n'}
\label{state}
	\end{eqnarray}
These states are not normalised. Their norm is easily computed; it is the square root of
\be
{\cal N}_j=\langle\psi_{j\vec n \vec n'} |\psi_{j\vec n \vec n'}\rangle =\sum_{k=-j}^j e^{-2\pi\gamma k}
\ee

These are the \BW link states. 

\subsection{Semiclassicality} \label{semi}

Before extending the \BW states to the full graph, let us study their properties. First of all, we have defined states at fixed spin $j$. Therefore we expect the corresponding conjugate momentum, namely the extrinsic curvature at the facet, to be fuzzy.  We leave open, for the moment, the task of combining these states into \emph{extrinsic}   \cite{Rovelli} semiclassical states, and we concentrate on the properties of the intrinsic geometry they define. 

For this, we estimate the mean value and the dispersion of the geometrical operators on the states  \eqref{state}. To begin with, consider the case with $\vec{n}=\vec{n}'=\vec z$. Choosing the basis that diagonalises $L_z$ we have immediately 
\begin{eqnarray} \label{lz}
{\langle \vec{L}_\su \rangle}&\equiv&	\langle \psi_{j\vec z \vec z} | \vec{L}_\su |\psi_{j\vec z \vec z} \rangle\nonumber \\ &=& \sum_{m} e^{-2\gamma \pi m} \langle jm| 	\begin{pmatrix} L_x \\ L_y \\ L_z  \end{pmatrix}  | jm \rangle\, , 
\end{eqnarray}
where, we recall,
\begin{equation*}
\begin{pmatrix} L_x \\ L_y \\ L_z  \end{pmatrix}  | jm' \rangle_\su=\begin{pmatrix} \frac{1}{2} (c_1 | j, m'-1 \rangle_\su  + c_2 | j, m'+1 \rangle_\su) \\ \frac{i}{2} (c_1 | j, m'-1 \rangle_\su  - c_2 | j, m'+1 \rangle_\su) \\ m'   | jm' \rangle_\su \end{pmatrix} 
\end{equation*}
the coefficients $c_1$, $c_2$ being defined by
\begingroup
    \addtolength{\jot}{.5 em}
	\begin{eqnarray*}
	c_1 &=& \langle j, m-1 | L_x -i L_y | j, m \rangle = \sqrt{(j+m) (j-m+1)}\\
c_2 &=& 	\langle j, m+1 | L_x + i L_y | j, m \rangle = \sqrt{(j+m+1) (j-m)}.
	\end{eqnarray*}
\endgroup
Easily, 
\be	
{\langle \vec{L}_\su \rangle}= \sum_{m= -j }^{j} e^{-2 \pi \gamma m} \begin{pmatrix} 0 \\ 0\\ m \end{pmatrix} = \left( \sum_{m= -j }^{j} e^{-2 \pi \gamma m} m \right) \, \vec{z}
\ee

The mean value, properly normalized, reads
\be
 \vec{L}_\su  \equiv \frac{	\langle  \vec{L}_\su \rangle }{{\cal N}_j} = 
 \frac{ \sum_{m= -j }^{j} e^{-2\pi \gamma m} m }{ \sum_{m= -j }^{j} e^{-2 \pi \gamma m}  }\ \vec{z}.
\ee
The vector operator points in the direction identified by the state and, for large $j$, we have:
\be
\frac{ \vec{L}_\su }{(-j)} \xrightarrow{j \rightarrow \infty} 1 \notag\\
\ee
Therefore, $\vec{L}_\su \sim -j + O(j)$. The correction is actually a constant given by:

\begin{equation}
 \vec{L}_\su \sim - j + \frac{1}{2(e^{\pi \gamma}-1)}
\end{equation} 
%\textcolor{red}{Comment about the Livine--Speziale case, scalar product.}
In order to understand if the state become sharp for large $j$, we look to the relative dispersion of $\vec{L}_\su$ in the plane orthogonal to the direction identified by the mean value. We saw above that
\begin{equation}
\langle L_x \rangle = \langle L_y \rangle = 0,
\end{equation}
while, due to symmetry, we can write
\be
\langle L_x^2 \rangle = \langle L_y^2 \rangle =\frac{1}{2} \langle (L_x^2 + L_y^2) \rangle 
=  \frac{1}{2} \langle (\vec{L}^2 - L_z^2) \rangle.
\ee
The relative dispersions of the components $L_x, L_y$ (the scale parameter is chosen to be the norm square of the vector itself) is given by
\begin{equation}
\frac{\sigma(L_x)}{\langle \vec{L}^2 \rangle} = \frac{\sigma(L_y)}{\langle \vec{L}^2 \rangle} =   \frac{1}{2} \frac{\langle (\vec{L}^2 - L_z^2) \rangle}{\langle \vec{L}^2 \rangle}.
\end{equation}
The expectation value of $\vec{L}^2 $ is $j (j+1)$, while for $ L_z^2$ we get
\be
 L_z^2  = 
%&\quad =\frac{\sum_{mm'n } e^{-\gamma \pi (m+ m')} \langle jm| L_z^2 | jm' \rangle_\su  }{ \langle \psi_j ^{(\vec{z}, \vec{n}')}  | \psi_j ^{ (\vec{z}, \vec{n}')}  \rangle } \delta_{mm'}^{\dagger}\\
\frac{ \sum_{m = -j}^j m^2 e^{-2 \gamma \pi m}}{ \sum_{m = -j}^j e^{-2 \gamma \pi m}}. \\
\ee

The spread is then given by~\footnote{The calculation is easy to perform if one notes that all the quantities appearing are all of the form:
\begin{equation}
\sum_{m= -j}^j m^n e^{- 2 \pi m} = \frac{d^n}{d\alpha^n} \sum_{m= -j}^j e^{- \alpha m}|_{\alpha= 2\pi}
\end{equation}
The only sum one needs to perform is then $\sum_{m= -j}^j e^{- \alpha m}$ which can be split into two geometric sums.}
\begin{eqnarray*}
\frac{\sigma(L_y)}{\langle \vec{L}^2 \rangle} = \frac{\sigma(L_x)}{\langle \vec{L}^2 \rangle}= \frac{1}{2} \frac{j(j+1)  - \frac{ \sum_{m = -j}^j m^2 e^{-2 \gamma \pi m}}{  \sum_{m = -j}^j e^{-2 \gamma \pi m}  }}{ j (j+1)},
\end{eqnarray*}
which goes to zero in the limit $j \to \infty$.

It is easy to generalize this result for a generic direction: the mean value of  $\vec{L}_\su$ on $|\psi_{j\vec{n}\vec{n}'}\rangle$ is given by
\begin{equation}
 \vec{L}_\su = \frac{1}{{\cal N}_j}\,\left( \sum_{m } e^{-2 \pi m} m \right)    \, \vec{n}.
\end{equation}
And the mean value of the right invariant vector field $\vec L_\ta$ is 
\begin{eqnarray}
\vec{L}_\ta= \frac{1}{{\cal N}_j} \,\left(\sum_n e^{- 2 \pi n}n\right) \, \vec{n}'.
\end{eqnarray}
with the same relative dispersion as above.

%So our states are not only labelled by classical value of the intrinsic geometry, given by the two normals $(\vec{n}, \vec{n}')$ and the quantum number $j$, both related to facet of the tetrahedrum dual to the link (and this define a mapping between classical and quantum framework); but they are also peacked on classical value of the variables describing the intrinsic geometry.

\subsection{Overcomplete basis}

Finally an important property of the \BW link states is that they form an overcomplete basis for each $j$, in the Hilbert space $\mathcal{H}_j \otimes \mathcal{H}^*_j$. The resolution of the identity is 
\begin{equation} \label{identity}
\mathbb{I}_j = \frac{d_j^2}{(4 \pi)^2 {\cal N}_j} \int_{S^2} d^2 \vec{n} \int_{S^2} d^2 \vec{n}' \ | \psi_{j\vec n \vec n'} \rangle \langle \psi_{j\vec n \vec n'} |
\end{equation}
The integration is over the two-sphere of the normalized vectors $\vec{n}, \vec{n}'$, with the standard $\mathbb{R}^3$ measure restricted to the unit sphere. This property is crucial: it indicates that every state can be expressed as a superposition of states with semiclassical labels. The  proof of \eqref{identity} is given in Appendix \ref{A}.\\

\section{Thermally correlated Spin network states}\label{sntherm}

So far we have studied single link states. We now move to states defined on the full graph. 
The first step for this is to combine \BW states associated to the links that join on a single node $n$.
To this aim, we simply take the tensor product of a \BW link state per each of the links meeting at $n$ and project on the $SU(2)$ gauge invariant subspace.
The projection is performed by integrating over the local gauge group $SU(2)$,
% making use of the invariant measure of the group, the Haar measure $dh$
\begin{eqnarray} \label{nodo}
{|\Psi^{(n)}_{j_l , \vec{n}_l,\vec{n}_l'} \rangle} =\int dh \bigotimes_{l \in n} D (h) |\psi_{j_l\vec n_l \vec {n}'_l}\rangle.
\end{eqnarray}

The \BW \emph{graph} state is then determined by a spin associated to each link and two vectors $\vec n_l$ and $\vec n_l'$ associated, respectively, to the source and the target of each link. The resulting gauge invariant state is
\be \label{spin-network}
 | \Psi_{j_l , \vec{n}_l,\vec{n}_l'} \rangle = \int \prod_{n} dh_{n} \bigotimes_{ l \equiv \langle n_l, n_l'\rangle}  D^{j_l}(h_{n_l}) D^{j_l \dagger}(h_{n'_l})|\psi_{j_l\vec n_l \vec n'_l}\rangle
\ee
where we identify each link with the two node at its endpoints, $l \equiv \langle n_l, n'_l\rangle$. 

These are the \BW states on the graph.

In the Schr\"odinger representation, namely on the group element basis, they read
\begin{eqnarray}
&& \Psi_{j_l , \vec{n}_l,\vec{n}_l'}  (U_l)  =\langle U_l |\Psi_{j_l , \vec{n}_l,\vec{n}_l'}  \rangle =  \int \prod_n dh_n \\ \nonumber 
&&\hspace{1em}    \prod_{ l \equiv \langle n_l, n_l'\rangle}       \tr_{j_l}[ D(U_l)D(h_{n_l}) D(\vec{n}_l) \,e^{- \pi L_z}\, D^{\dagger} (\vec{n}'_l) D^{\dagger}(h_{n'_l})].
\end{eqnarray}

These states resemble the common intrinsic Livine-Speziale states on the graph, but there is a crucial difference. The space of the states with fixed spin is the tensor product of one intertwined space per node, that is
\be
{\cal H}_{j_l}=\bigotimes_n {\cal H}_n
\ee
where the intertwined space ${\cal H}_n$ of the node $n$ is the $SU(2)$ invariant part of the tensor product of the representation spaces associated to the spins of the links joining in $n$:
\begin{equation*}
{\cal H}_n=Inv_{SU(2)}[\bigotimes_l \mathcal{H}_{j_l}] .
\end{equation*}
where the product in $l$ runs over the links joining in $n$.  The Livine-Speziale states  $|j_l , \vec{n}_l,\vec{n}_l'\rangle$ are tensor states with respect to this decomposition
\be
 | j_l , \vec{n}_l,\vec{n}_l'\rangle = \bigotimes_n  | \iota^n_{j_l , \vec{n}_l,\vec{n}_l'}\rangle 
\ee
where $\iota^n_{j_l , \vec{n}_l,\vec{n}_l'}$ is the Livine-Speziale intertwiner.  On the contrary, the \BW states do not factorise.  To see this, it is sufficient to consider the density matrix of the state defined in (\ref{nodo}) and reduce it to the intertwined space ${\cal H}_n$, by tracing over the external representation spaces of the links. A straightforward calculation shows this to be 

\begin{eqnarray*}
&&\rho  =  \tr [ |\Psi^{(n)}_{j_l , \vec{n}_l,\vec{n}_l'}\rangle \langle  \Psi^{(n)}_{j_l , \vec{n}_l,\vec{n}_l'}|]= \\
&\quad&= \bigotimes_{l\in n} \int dh \int d\tilde{h} \quad [D(h) e^{- 2 \pi \gamma \vec{L}_l \cdot \vec{n}_l} D^{\dagger} (\tilde{h})]\\
\end{eqnarray*}
where the tensor product is on the links that join at the node $n$ and for simplicity we have assumed the node to be the source of these all.  One may notice in this expression that $[D(h) e^{- 2 \pi \sum_l \vec{L}_l \cdot \vec{n}_l} D^{\dagger} (\tilde{h})] $ does not act as a rotation of the vectors $\vec{n}_l$, since the adjoint rapresentation acts with the same group element, whereas here we have two different $SU(2)$ elements ($h, \tilde{h}$). This density matrix in general is not pure.

This indicates that the \BW states carry nontrivial quantum correlations between different nodes. In Appendix \ref{dip} we compute the correlations between two operators in a \BW state on a simple graph (the dipole graph), to  verify explicitly that they are indeed non-vanishing (see also Appendix \ref{dipo} for details on the observables).

This is the main property we were seeking. 

\section{Long distance correlations} \label{propa}

The \BW states defined in the previous section have non trivial quantum correlations across adjacent nodes.  Do they also have correlations between nodes that are not adjacent?  Here we show that the answer is yes and we give some preliminary elements of analysis of these correlations. 

The simplest spin network we can use to try to address these questions, is an open spin network  composed by a chain of N nodes, each pair sharing a single link.  We start by writing the explicit form of the state for the special case $N=2$ to understand the structure of the state itself. The non-gauge invariant state on the two node graph is given by
\begin{eqnarray*}
&&| \tilde{\Psi}_{j_l , \vec{n}_l , \vec{n}'_l}\rangle = \sum_{ \{k_l\} } e^{- \pi \gamma \sum_l k_l} |j_{1234}, k_{1234}, \vec{n}_{1234} \rangle \times \\
&& \qquad \times \,| j_{4567}, k_{4567} \vec{n}'_{4567} \rangle^{\dagger} ||j_l , k_l, \vec{n}^{(')}_{l} \rangle ^{(\dagger)}_{i \neq 4}
\end{eqnarray*}
gathering together the external half links into the espression $||j_l , k_l, \vec{n}^{(')}_{l} \rangle ^{(\dagger)}_{l \neq 4}$.
The projection to the gauge invariant subspace is given by
\begingroup
    \addtolength{\jot}{.5 em}
	\begin{align*}
&| \Psi_{ j_l , \vec{n}_l , \vec{n}'_l} \rangle = \\
&\qquad = \sum_{ \{k_l\} } e^{- \pi \gamma \sum_l k_l} \sum_{\alpha \beta} \phi_{\alpha}(j_{1234} k_{1234}, \vec{n}_{1234}) \times \\
& \qquad \times \,  \phi^{*}_{\beta}( j_{4567}, k_{4567} \vec{n}'_{4567} )\, | \iota_{\alpha} \rangle | \iota_{\beta} \rangle^{\dagger} ||j_l , k_l, \vec{n}^{(')}_{l} \rangle ^{(\dagger)}_{l \neq 4}
\end{align*}
\endgroup
where the projector operator and the $\phi_{\alpha, \beta}$ coefficients are those defined in Appendix \ref{dip}.
The generalization to a chain of $N$ nodes is straightforward. 

\begin{figure}[h]
\includegraphics[width=7cm]{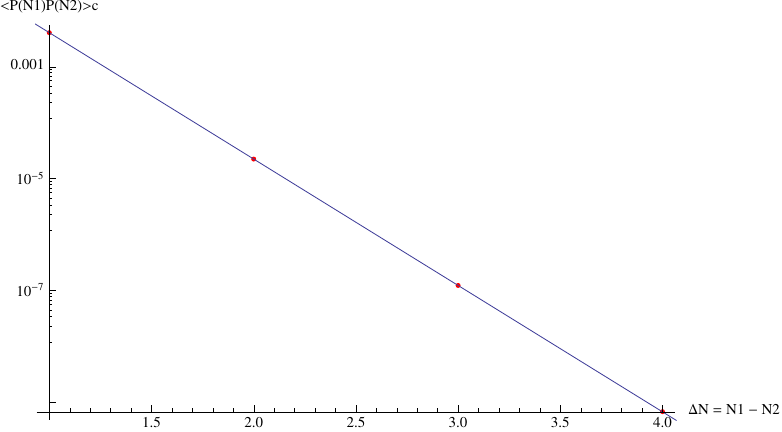}
 \caption{Fit of the correlation function ($\langle P(N1) P(N2) \rangle_c = \langle P(N1) P(N2) \rangle -\langle P(N1) \rangle\langle P(N2) \rangle $ ) as a function of the distance between nodes ($\Delta N = N1 - N2$). The scale is linear logarithmic.  Fit model: $f(\Delta N) =  a \,\exp( - b \Delta N)$.
Fit results: $ a = 0, 73 ; b = 5,20 $. }\label{fig2}
 \end{figure}

We have computed numerically the correlations on a chain of $N=7$ nodes, fixing all $j_l = 1/2$. We have computed the following quantity
\begin{equation}
\langle P_i^{(0)} P_j^{(0)} \rangle - \langle P_i^{(0)} \rangle\langle P_j^{(0)} \rangle
\end{equation}
for $i, j= 1, \cdots , N$, where $P_i^{(0)}= |\iota_0 \rangle \langle \iota_0|$ is the projector on the first element of the recoupling basis on the  $i$-th node. The results of the numerical calculation are displayed in Figure \ref{fig2}.

The correlations that we find can be interpreted as the result  the interplay between the thermal correlations on the single links, which correlate any two adjacent nodes, and the effect of gauge-invariance at nodes, which ties the links in quadruples and allows for the propagation of the those thermal correlations among far nodes. 

\section{Summary and Discussion}

We have defined a family of states in loop quantum gravity which are peaked on an (intrinsic) geometry and have non trivial correlations between distinct nodes.  These correlations are such that tracing on one node yields, on a neighbouring node, a thermal state with respect to a flow related, via the simplicity conditions, to the boost generator. We have called these states \BW states, and this feature the \BW property. Correlations extend to non-neighbouring nodes.

We list in the following a number of questions which we think deserve to be investigated. 

\begin{itemize}

\item We have investigated in this paper states at fixed $j$.  Extrinsic coherent states obtained  relaxing the sharpness condition on $j$ are of course  interesting for physics. 

\item The boost operator $\vec K$ is the generator of \emph{internal} boosts in the full covariant theory.  In a gauge fixed formalism, as is implicitly the loop formalism which is formulated in the time gauge, this is related to a physical boost.  The situation is analogous to the rotations of the tetrad in general relativity: if we describe a measuring apparatus gauge fixing the tetrad to its axes, then a rotation of the tetrad has the physical interpretation of a relative rotation between the apparatus and its exterior.  Similarly, the boost generator $\vec K$ can also be interpreted as the generator of \emph{physical} boosts: it evolves a state on a surface to the state on a boosted surface, which is to say to the surfaces of a boosted observer \cite{Rovelli:1990ph}.  

\item In the context of QFT, the restriction of the field vacuum state to the right Rindler wedge, automatically gives a restriction to the positive eigenvalues of the boost generator. In closer analogy, perhaps the restriction of states to those where $\vec{K} \cdot \vec{n}$ has positive eigenvalues is of physical interest. In particular, this aspect could turn out to be crucial for the normalizability of the states once the sum over the $SU(2)$ representations ($j$) is considered. We leave this question open.

\item The spin-networks Hilbert space is the same as in a conventional $SU(2)$ lattice Yang-Mills theory. In lattice gauge theory, nonlocal correlations have been studied by Donnelly in \cite{Donnelly2012a}. See also \cite{Buividovich2008}. The construction here is related to these analyses, and the precise relation deserves to be better understood. 

\item In the \BW states, the thermal correlations get mixed up by gauge invariance. The density matrix on a single node Hilbert space obtained by tracing the state on the rest of the nodes is not thermal, because the correlations defined on the links get mixed up by the gauge at the node.  This is not in contradiction with the \BW property, which refers to a single surface (a single link), but deserves better understanding.

\item The $2\pi$ in the \BW temperature is related to the Minkowski geometry and its complex extension, as well as the corner terms of the action on the spitting surface \cite{Hayward:1993my,Neiman:2011gf,Bianchi:2012vp}. The role of these in the loop dynamics is strictly connected to the \BW property and is a tantalising issue, which still deserves clarification. 

\end{itemize}

\section*{acknowledgements}

Thanks to Aldo Riello and Hal Haggard for many exchanges. PR thanks the support of the Della Riccia Foundation.

\vspace{4em}

\appendix 

\section{Resolution of identity}\label{A}
Here we derive explicitly the resolution of the identity given in the text. Starting from equation  \eqref{identity}, we have
\begin{eqnarray}
\mathbb{I}_j &=& \frac{1}{\mathcal{N}_j}  \frac{d_j^2}{(4 \pi)^2}  \int_{S^2} d^2 \vec{n} \int_{S^2} d^2 \vec{n}' | \psi_j^{(\vec{n}, \vec{n}')} \rangle \langle \psi_j^{(\vec{n}, \vec{n}')} | =\\ \nonumber
&=& \frac{1}{\mathcal{N}_j}  \frac{d_j^2}{(4 \pi)^2} \int_{S^2} d^2 \vec{n} \int_{S^2} d^2 \vec{n}'    \sum_k |j, k \rangle_\su D(\vec{n}) \, e^{- \pi \gamma k}  \times\\  \nonumber
&& \times\,\,  D^{\dagger} (\vec{n}') |j, k \rangle_\su^{\dagger} \sum_l |j, l \rangle_\ta D(\vec{n}') e^{- \pi \gamma l} D^{\dagger} (\vec{n}) |j, l \rangle_\ta^{\dagger}  \\ \nonumber
&=& \frac{1}{\mathcal{N}_j}   \frac{d_j^2}{(4 \pi)^2} \int_{S^2} d^2 \vec{n} \int_{S^2} d^2 \vec{n}'    \sum_{k, \alpha, \beta} |j, \alpha \rangle_\su D(\vec{n})_{\alpha, k} e^{- \pi \gamma k} \\  \nonumber
&&\hspace{-1em}  D^{\dagger} (\vec{n}')_{k, \beta}    |j, \beta \rangle_\ta^{\dagger}  \sum_{l, \tilde{\alpha}, \tilde{\beta}} |j, \tilde{\beta} \rangle_\ta D(\vec{n}')_{\tilde{\beta}, l} e^{- \pi \gamma l} D^{\dagger} (\vec{n})_{l, \tilde{\alpha}} |j, \tilde{\alpha} \rangle_\su^{\dagger}. 
\end{eqnarray}
Rearranging factors,
\begin{eqnarray}  \nonumber
\mathbb{I}_j &=& \frac{1}{\mathcal{N}_j}  d_j^2 \sum_{k, \alpha, \beta}    \sum_{l, \tilde{\alpha}, \tilde{\beta}} |j, \alpha \rangle_\su   \langle j, \tilde{\alpha} |_\su    |j, \tilde{\beta} \rangle_\ta \langle  j, \beta |_\ta  e^{- \pi \gamma (k+l)} \\            \nonumber
&&  \int_{S^2} \frac{d^2 \vec{n}}{4 \pi}     D(\vec{n})_{\alpha, k} D^{\dagger} (\vec{n})_{l, \tilde{\alpha}} 
        \int_{S^2} \frac{d^2 \vec{n}'}{4 \pi}       D^{\dagger}(\vec{n}')_{k, \beta}  D(\vec{n}')_{\tilde{\beta}, l}\\  \nonumber
&=& \frac{1}{\mathcal{N}_j}  d_j^2 \sum_{k, \alpha, \beta}    \sum_{l, \tilde{\alpha}, \tilde{\beta}} 
          |j, \alpha \rangle_\su   \langle j, \tilde{\alpha} |_\su    |j, \tilde{\beta} \rangle_\ta \langle  j, \beta |_\ta \times \\   \nonumber
&&\times\,\, e^{- \pi \gamma (k+l)} \frac{  \delta_{\alpha, \tilde{\alpha}} \delta_{k, l}  }{d_j}  \frac{\delta_{\beta, \tilde{\beta}} \delta_{k, l}  }{d_j} \\  \nonumber
&=& \frac{1}{\mathcal{N}_j}  \sum_{k, \alpha, \beta} |j, \alpha \rangle \langle j, \alpha |  |j, \beta \rangle \langle j, \beta | e^{- 2 \pi k} \\  \nonumber
&=& \sum_{\alpha} |j, \alpha \rangle \langle j, \alpha | \times \sum_{ \beta} |j, \beta \rangle \langle j, \beta | \times \frac{ \sum_{k} e^{- 2 \pi k} }{ \sum_{k} e^{- 2 \pi k} }  \\
&=&  \mathbb{I}_\su \otimes \mathbb{I}_\ta
\end{eqnarray}
which is the identity in $\mathcal{H}_j \otimes \mathcal{H}^*_j$.

\section{Node observables} \label{dipo}
\begin{figure}[h]
\includegraphics[width=6cm]{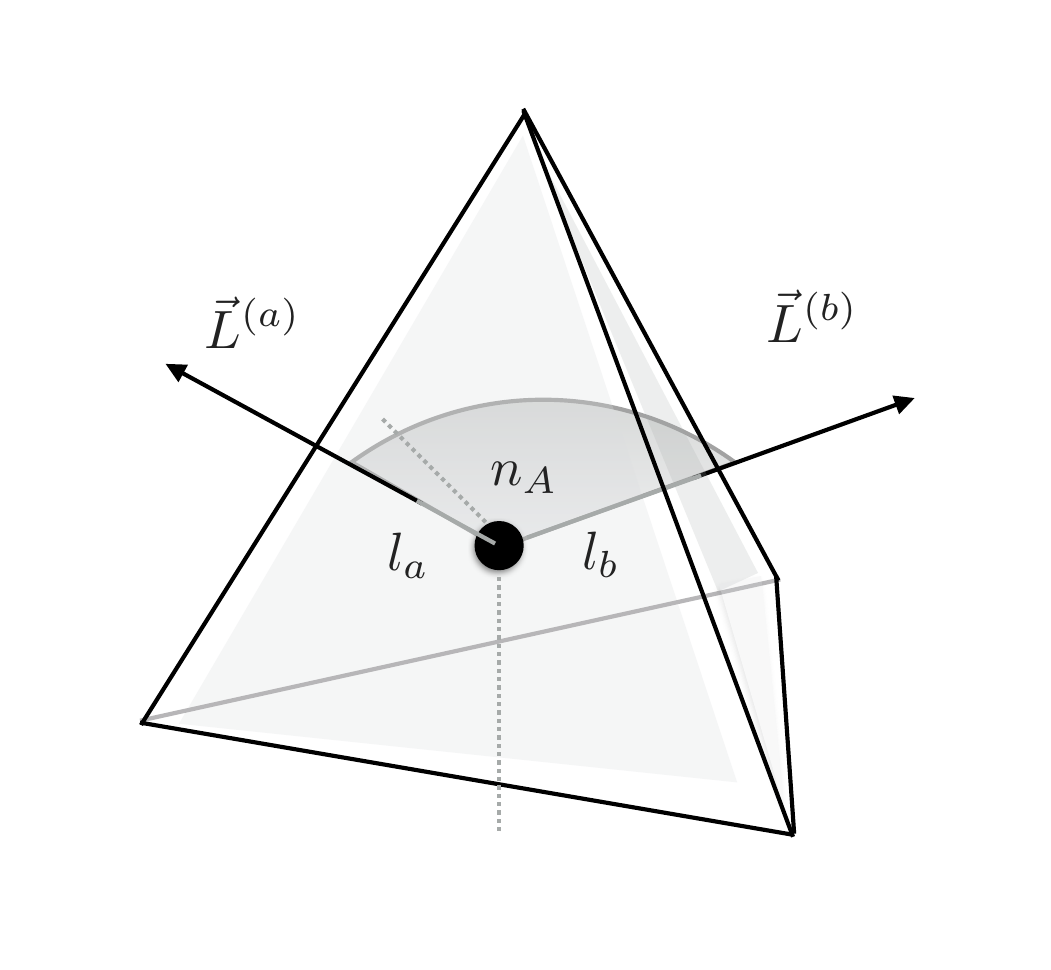}
\caption{Thetrahedron cell dual to the node A. The two links $l_a$ and $l_b$ meet at the node $A$. The observable given by the scalar product $(\vec{L}^{(a)} \cdot \vec{L}^{(b)})_A$ is a measure of the dihedral angle (shaded above) between the two facets of the cell $A$.}\label{dihe}
\end{figure}
To study correlations we need an observable to probe them.   A good example of observable is the scalar product $(\vec{L}^{(a)} \cdot \vec{L}^{(b)})_A$ (where $a$ and $b$ are two links that meet at the node $A$). Its geometrical interpretation is a measure of the dihedral angle between the two facets of the cell $A$. Recall that the area of these facets is $|L_a|$ and $|L_b|$.   Here we assume for simplicity that all nodes are four-valent. See Figure \ref{dihe}. 

This observable is diagonal in the appropriate recoupling basis, $ | \iota_{\alpha} \rangle \equiv \{| j_1 \cdots j_4, \iota_{\alpha} \rangle \} $, labelled by the spin number $\alpha$ of the ``virtual link'' associated to the node \cite{Rovelli}. This can be seen explicitly by looking at the operator $(\vec{L}^{(a)} + \vec{L}^{(b)})^2$ first: consider the explicit form of the intertwiner state
\begin{eqnarray} \label{iota}
| \iota_{\alpha} \rangle &=& | j_1 \cdots j_4, \iota_{\alpha} \rangle =\\ \nonumber
 & = & \sum_{k_1, k_2, k_3, k_4} \iota^{k_1 k_2 k_3 k_4}_{\alpha} |j_1 k_1 \rangle |j_2 k_2 \rangle |j_3 k_3 \rangle |j_4 k_4 \rangle \\ \nonumber
&=&  \sum_{\substack{k_1, k_2, k_3, k_4\\ m}} \iota^{k_1 k_2 m} \iota_{m}^{k_3 k_4} |j_1 k_1 \rangle |j_2 k_2 \rangle |j_3 k_3 \rangle |j_4 k_4 \rangle .
\end{eqnarray}
Acting with the operator $(\vec{L}^{(a)} + \vec{L}^{(b)})$, where $a=1, b=2$ we have 
\be
(L^{i\,(1)} + L^{i\,(2)}) | j_1 k_1 \rangle |j_2 k_2 \rangle =  [J^{i(j_1)}  + J^{i(j_2)}] |j_1 k_1 \rangle |j_2 k_2 \rangle
\ee
where $J^i$ are the generators of $SU(2)$ in the rapresentation $j$.
If we now use this for the full state (\ref{iota}), we have
\begin{eqnarray}
&&(L^{i\,(1)} + L^{i\,(2)}) | \iota_{\alpha} \rangle= \\ 
&=&\sum_{\substack{k_1, k_2, k_3, k_4 \\ m}} {\iota^{k_1 k_2}}_m \iota^{m k_3 k_4} [J^{i(j_1)}  + J^{i(j_2)}] \times\\ \nonumber
&& \quad \times\, |j_1 k_1 \rangle |j_2 k_2 \rangle |j_3 k_3 \rangle |j_4 k_4 \rangle\\ \nonumber
&=& \sum_{\substack{k_1, k_2, k_3, k_4 \\ m}} {\iota^{k_1 k_2}}_{ m} [- J^{i(\alpha)}]  \iota^{m k_3 k_4} |j_1 k_1 \rangle |j_2 k_2 \rangle |j_3 k_3 \rangle |j_4 k_4 \rangle \\ \nonumber
&=& \sum_{\substack{k_1, k_2, k_3, k_4 \\ m, \tilde{m} }} \iota^{k_1 k_2 \tilde{m}} [- J^{(\alpha)}]^{i}_{\tilde{m} m}  \iota^{ m k_3 k_4} |j_1 k_1 \rangle |j_2 k_2 \rangle |j_3 k_3 \rangle |j_4 k_4 \rangle 
\end{eqnarray}
where we used the definition of the intertwiner, $ D^{(j_1)}_{m_1 n_1} D^{(j_2)}_{m_2 n_2} D^{(j_1)}_{m_3 n_3} \iota^{n_1 n_2 n_3} = \iota^{m_1 m_2 m_3}$ to get
\begin{eqnarray*}
&\quad & ( J^{(j_1)}+ J^{(j_2)} + J^{(j_3)})\, \iota^{n_1 n_2 n_3} =0 \\
& \Rightarrow &  ( J^{(j_1)}+ J^{(j_2)} )\, \iota^{n_1 n_2 n_3} =  - J^{(j_3)} \iota^{n_1 n_2 n_3}
\end{eqnarray*}
Finally, applying the same operator a second time, we obtain
\begin{eqnarray*}
&&(L^{i\,(1)} + L^{i\,(2)}) (L^{i\,(1)} + L^{i\,(2)}) | \iota_{\alpha} \rangle  = \\
&&= \sum_{\substack{k_1, k_2, k_3, k_4 \\ m}} {\iota^{k_1 k_2}}_m [- J^{i(\alpha)}]^2  \iota^{m k_3 k_4} |j_1 k_1 \rangle |j_2 k_2 \rangle |j_3 k_3 \rangle |j_4 k_4 \rangle \\
&&=  \alpha (\alpha +1) \sum_{\substack{k_1, k_2,  k_3, k_4 \\ m}} {\iota^{k_1 k_2}}_m   \iota^{m k_3 k_4} |j_1 k_1 \rangle |j_2 k_2 \rangle |j_3 k_3 \rangle |j_4 k_4 \rangle \\
&&=   \alpha (\alpha +1)  | \iota_{\alpha} \rangle 
\end{eqnarray*}
since $[J^{i(\alpha)}]^2$ is the Casimir operator.
Analogously, the operator $(L^{i\, (a)} L^{i\,(b)})$ will be diagonal on this basis, as
\begin{equation}
L^{i\,(1)}\, L^{i\,(2)} = \frac{1}{2} [ (L^{i\,(1)} + L^{i\,(2)})^2 - (L^{i\,(1)})^2 - (L^{i\,(2)})^2]
\end{equation}
with eigenvalues given by
\begin{equation}
C_{\alpha}= \alpha (\alpha + 1) - j_1 (j_1 + 1) - j_2 (j_2 +1).
\end{equation}
In the recoupling basis, the operator takes the form
\begin{equation}
L^{i\,(1)} L^{i\,(2)}  = \sum_{\alpha} C_{\alpha} |\iota_{\alpha} \rangle \langle \iota_{\alpha}|.
\end{equation}

\section{Correlation in dipole graph}\label{dip}

\begin{figure}[h]
 \centering
   {\includegraphics[width=7cm]{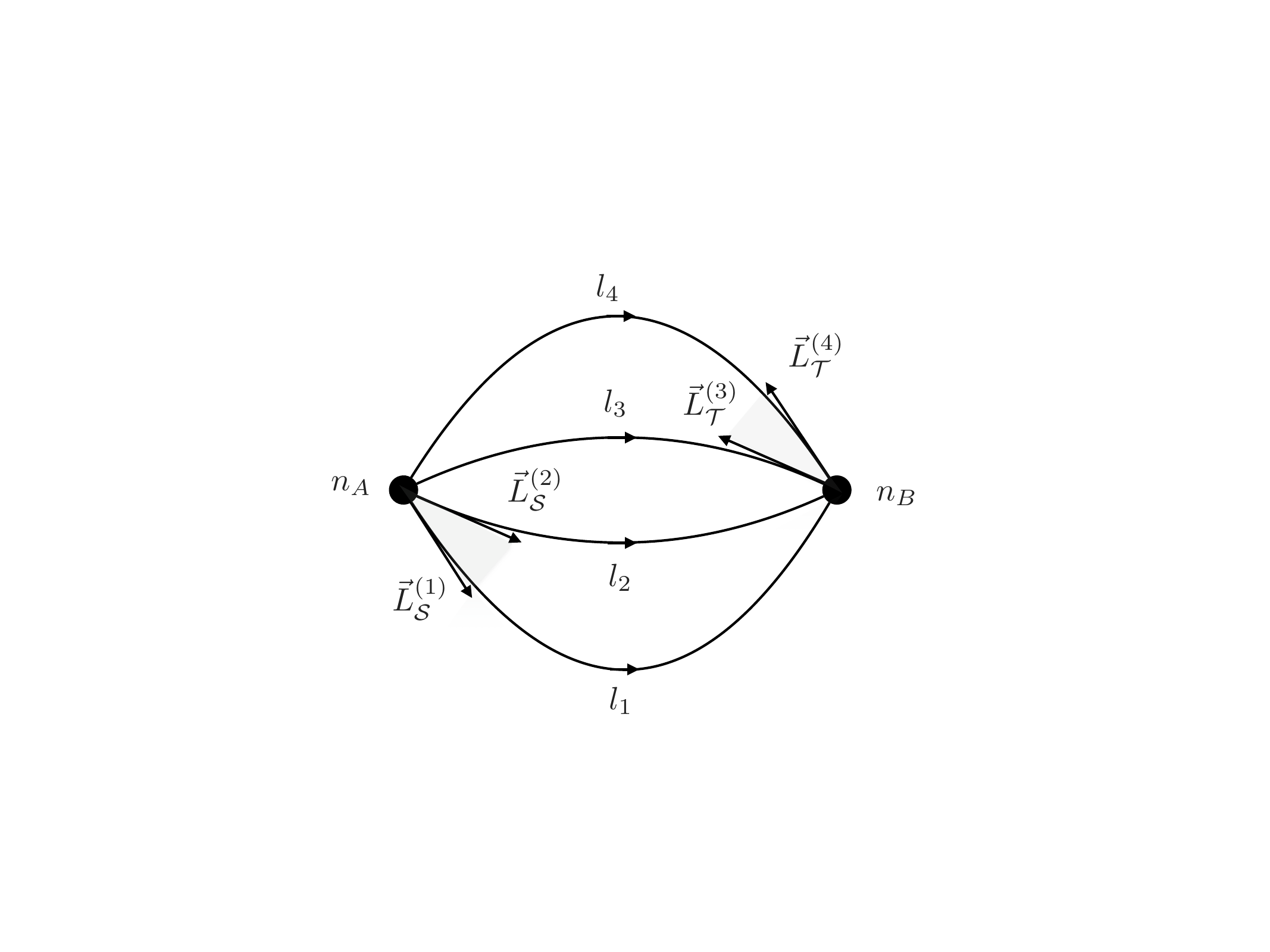}}
 \caption{Dipole graph $\Delta^*$: two four-valent nodes, $A$ and $B$, sharing four links.  We consider two operators acting on the two nodes: $(\vec{L}_{\su}^{(1)} \cdot \vec{L}_{\su}^{(2)})_A$ and $(\vec{L}_{\ta}^{(3)} \cdot \vec{L}_{\ta}^{(4)})_B$}\label{dipolo}
 \end{figure}

Here we show that the correlations between nodes are in fact non vanishing, by providing a detailed example for a simple graph. We consider the \textit{dipole} graph $\Delta^*$: two four-valent nodes, $A$ and $B$, sharing four links (Figure \ref{dipolo}).  We consider two operators acting on the two nodes: $(\vec{L}_{\su}^{(1)} \cdot \vec{L}_{\su}^{(2)})_A$ and $(\vec{L}_{\ta}^{(3)} \cdot \vec{L}_{\ta}^{(4)})_B$, with $\vec{L}_{\su}$ and $\vec{L}_{\ta}$ being respectively the left and right invariant vector fields.
We want to measure the correlation between the two nodes.

We expand the state in the appropriate recoupling basis, where the chosen operators are diagonal:
\begin{itemize}
\item {$  |\iota_{\alpha} \rangle \equiv |j_1, j_2, j_3, j_4 \iota_{\alpha} \rangle$  s.t. \quad$\iota_{\alpha}^{k1 k_2 k_3 k_4} = \iota^{k_1 k_2 a} \iota_a^{k_3 k_4}  $}\\
\item $| \iota_{\beta} \rangle \equiv |j_1, j_2, j_3, j_4 \iota_{\beta} \rangle$   s.t. \quad  $\iota_{\beta}^{k_1 k_2 k_3 k_4} = \iota^{k_1 k_2 b} \iota_b^{k_3 k_4}$ 
 \end{itemize}
Instead of integrating over the group for each node, we impose the gauge invariance through the projectors
\begin{equation} \label{projectors}
P_A = \sum_{\alpha} | \iota_{\alpha} \rangle \langle \iota_{\alpha}| \quad \text{and}\quad P_B = \sum_{\beta} | \iota_{\beta} \rangle \langle \iota_{\beta} |. 
\end{equation}
The non-gauge invariant state is given by
\begin{eqnarray*}
&&| \tilde{\Psi}_{j_l , \vec{n}_l , \vec{n}'_l} \rangle =\\
&\quad&= \sum_{ \{k_l\} } e^{- \pi \gamma \sum_l k_l} |j_{1234}, k_{1234}, \vec{n}_{1234} \rangle_\su  | j_{1234}, k_{1234} \vec{n}'_{1234} \rangle^{\dagger}_\ta 
\end{eqnarray*}

Projecting  with (\ref{projectors}), we get its gauge invariant version
\begin{eqnarray*}
&&| \Psi_{ j_l , \vec{n}_l , \vec{n}'_l } \rangle = \sum_{ \{k_l\} } e^{- \pi \gamma \sum_l k_l}\times\\ 
&& \times \,\sum_{\alpha \beta} \phi_{\alpha}(j_{1234} k_{1234}, \vec{n}_{1234}) 
 \phi^{*}_{\beta}( j_{1234}, k_{1234} \vec{n}'_{1234} ) | \iota_{\alpha} \rangle | \iota_{\beta} \rangle^{\dagger} 
\end{eqnarray*}
where we used the definition: 
\begin{eqnarray}\label{fi}
 \phi_{\alpha}(j_{1234} k_{1234}, \vec{n}_{1234}) &=& \langle \iota_{\alpha} |j_{1234}, k_{1234}, \vec{n}_{1234} \rangle  \\
 \phi_{\beta}( j_{1234}, k_{1234} \vec{n}'_{1234} ) &=& \langle \iota_{\beta} |  j_{1234}, k_{1234}, \vec{n}'_{1234} \rangle \nonumber
\end{eqnarray}
We can write these coefficients explicitly. Consider first the case in which $\vec{n}_i  \equiv \vec{z}$.
In this case we have
\begin{equation}
 \langle \iota_{\alpha} |j_{1234}, k_{1234} \rangle = \iota^{k_1 k_2 k_3 k_4 }_{\alpha} 
\end{equation}
whose components can be calculated using the decomposition of this invariant tensor with $\{ 3j \}$ symbols.
If instead we keep generic directions $\vec{n}_i$, we need to take into account a rotation matrix for each link:
\begin{eqnarray*}
&&\langle \iota_{\alpha} |j_{1234}, k_{1234}, \vec{n}_{1234} \rangle = \sum D^{j_1}_{l_1 k_1} (\vec{n}_1) D^{j_2}_{l_2 k_2} (\vec{n}_2)\times \\
&& \qquad \times \, D^{j_3}_{l_3 k_3} (\vec{n}_3)D^{j_4}_{l_4 k_4} (\vec{n}_4)\, \iota^{l_1 l_2 l_3 l_4 }_{\alpha} 
\end{eqnarray*}
Now we can take the expectation value of the operators. We obtain
\begin{eqnarray*}
&&\langle \Psi_{ j_l , \vec{n}_l , \vec{n}'_l } | (\vec{L}_{\su}^{(1)} \cdot \vec{L}_{\su}^{(2)})_A (\vec{L}_{\ta}^{(3)} \cdot \vec{L}_{\ta}^{(4)})_B | \Psi_{ j_l , \vec{n}_l , \vec{n}'_l } \rangle = \\
&\quad&= \sum_{ \{k_l \} , \{ \tilde{k}_l \}}e^{- \pi \gamma \sum_l (k_l +\tilde{k}_l)} \times \\
&& \quad \times \,\sum_{\alpha, \tilde{\alpha}} \phi^{\{k_l , \vec{n}_l \} }_{\alpha} \phi^{*\{\tilde{k}_l\, \vec{n}_l \} }_{\tilde{\alpha}}  \langle \iota_{\tilde{\alpha}} |  (\vec{L}_{\su}^{(1)} \cdot \vec{L}_{\su}^{(2)})_A | \iota_{\alpha} \rangle \times \\
&& \quad \times \, \sum_{\beta, \tilde{\beta}}  \phi^{*\{k_l\, \vec{n}'_l \} }_{\beta} \phi^{\{\tilde{k}_l, \vec{n}'_l \}}_{\tilde{\beta}} \langle \iota_{\tilde{\beta }} | (\vec{L}_{\ta}^{(3)} \cdot \vec{L}_{\ta}^{(4)})_B | \iota_{\beta} \rangle
\end{eqnarray*}
\begin{eqnarray*}
&\quad &= \sum_{ \{k_l \} , \{ \tilde{k}_l \} }  e^{- \pi \gamma \sum_l (k_l +\tilde{k}_l)} \sum_{\alpha, \tilde{\alpha}} \phi^{ \{ k_l , \vec{n}_l \} }_{\alpha} \phi^{* \{  \tilde{k}_l, \vec{n}_l \} }_{\tilde{\alpha}}   C_{\alpha} \delta_{\alpha, \tilde{\alpha}}  \times \\
&& \quad \times \, \sum_{\beta, \tilde{\beta}}  \phi^{*\{ k_l , \vec{n}'_l \} }_{\beta} \phi^{\{\tilde{k}_l , \vec{n}'_l \}}_{\tilde{\beta}} C_{\beta} \delta_{\beta, \tilde{\beta}}\\
&\quad &= \sum_{ \{k_l \} , \{ \tilde{k}_l \} }  e^{- \pi \gamma \sum_l (k_l +\tilde{k}_l)} \sum_{\alpha, \tilde{\alpha}} \phi^{ \{ k_l , \vec{n}_l \} }_{\alpha} \phi^{* \{  \tilde{k}_l, \vec{n}_l \} }_{\alpha}   C_{\alpha}  \times \\
&& \quad \times \, \sum_{\beta, \tilde{\beta}}  \phi^{*\{ k_l , \vec{n}'_l \} }_{\beta} \phi^{\{\tilde{k}_l , \vec{n}'_l \}}_{\beta} C_{\beta} 
\end{eqnarray*}

Similarly,
\begin{eqnarray*}
&&\langle \Psi_{ j_l , \vec{n}_l , \vec{n}'_l } | (\vec{L}_{\su}^{(1)} \cdot \vec{L}_{\su}^{(2)})_A | \Psi_{ j_l , \vec{n}_l , \vec{n}'_l } \rangle = \\
&\quad & = \sum_{ \{k_l \} , \{ \tilde{k}_l \} }  e^{- \pi\gamma \sum_l (k_l +\tilde{k}_l)}  \sum_{\alpha, \tilde{\alpha}} \phi^{ \{ k_l , \vec{n}_l \} }_{\alpha} \phi^{* \{  \tilde{k}_l, \vec{n}_l \} }_{\alpha}   C_{\alpha} \times \\
&& \quad \times \sum_{\beta, \tilde{\beta}}  \phi^{*\{ k_l , \vec{n}'_l \} }_{\beta} \phi^{\{\tilde{k}_l , \vec{n}'_l \}}_{\beta} 
\end{eqnarray*}
and
\begin{eqnarray*}
&&\langle \Psi_{ j_l , \vec{n}_l , \vec{n}'_l } | (\vec{L}_{\ta}^{(3)} \cdot \vec{L}_{\ta}^{(4)})_B | \Psi_{ j_l , \vec{n}_l , \vec{n}'_l } \rangle =\\
& \quad & = \sum_{ \{k_l \} , \{ \tilde{k}_l \} }  e^{- \pi \gamma \sum_l (k_l +\tilde{k}_l)}  \sum_{\alpha, \tilde{\alpha}} \phi^{ \{ k_l , \vec{n}_l \} }_{\alpha} \phi^{* \{  \tilde{k}_l, \vec{n}_l \} }_{\alpha} \times \\
&& \quad \times \sum_{\beta, \tilde{\beta}}  \phi^{*\{ k_l , \vec{n}'_l \} }_{\beta} \phi^{\{\tilde{k}_l , \vec{n}'_l \}}_{\beta} C_{\beta} .\\
\end{eqnarray*}
We want to prove the following inequivalence
\begin{eqnarray*}
&\langle (\vec{L}_{\su}^{(1)} \cdot \vec{L}_{\su}^{(2)})_A (\vec{L}_{\ta}^{(3)} \cdot \vec{L}_{\ta}^{(4)})_B  \rangle & \\
& \mathrel{\rotatebox{270}{$\neq$}} & \\
&\langle  (\vec{L}_{\su}^{(1)} \cdot \vec{L}_{\su}^{(2)})_A \rangle  \langle  (\vec{L}_{\ta}^{(3)} \cdot \vec{L}_{\ta}^{(4)})_B \rangle & .
\end{eqnarray*}
Since we know the explicit form of the coefficients, we can verify this statement in an explicit example. For semplicity, let us fix all $j_l = 1/2$ on the links, so that the intertwiner basis (recoupling basis) has only two elements $\{ |\iota_0 \rangle , |\iota_1 \rangle\}$, and consider $\vec{n}_l = \vec{n}'_l = \vec{z}$, for each $l$.
We find
\begingroup
\addtolength{\jot}{.5 em}
\begin{align} 
\langle  (\vec{L}_{\su}^{(1)} \cdot \vec{L}_{\su}^{(2)})_A (\vec{L}_{\ta}^{(3)} \cdot \vec{L}_{\ta}^{(4)})_B   \rangle&=\,\frac{5}{4}& \\
\langle  (\vec{L}_{\su}^{(1)} \cdot \vec{L}_{\su}^{(2)})_A  \rangle =  \langle  (\vec{L}_{\ta}^{(3)} \cdot \vec{L}_{\ta}^{(4)})_B  \rangle &=-\frac{1}{2}& 
	\end{align}
\endgroup
which implies
\begingroup
\addtolength{\jot}{.5 em}
\begin{eqnarray} 
&&\langle (\vec{L}_{\su}^{(1)} \cdot \vec{L}_{\su}^{(2)})_A (\vec{L}_{\ta}^{(3)} \cdot \vec{L}_{\ta}^{(4)})_B   \rangle - \\
&&\qquad - \langle  (\vec{L}_{\su}^{(1)} \cdot \vec{L}_{\su}^{(2)})_A \rangle  \langle (\vec{L}_{\ta}^{(3)} \cdot \vec{L}_{\ta}^{(4)})_B \rangle =  \frac{3}{4} \neq 0 \nonumber
	\end{eqnarray}
\endgroup
The conclusion is that the \BW states have correlations between neighbouring nodes.

%\bibliographystyle{utcaps} 
%\bibliography{library}

\end{document}